# Electric Field Induced Topological Phase Transition in Two-Dimensional Few-layer Black Phosphorus

Qihang Liu[1,*], Xiuwen Zhang[1], L. B. Abdalla[2], A. Fazzio[2] and Alex Zunger[1,*]

[1]University of Colorado, Boulder, CO 80309, USA

[2]Instituto de Física, Universidade de São Paulo, SP, Brazil

[*]E-mail: qihang.liu85@gmail.com; alex.zunger@gmail.com.

Topological insulators (TI) have exciting physical properties but unfortunately there are extremely few known TI's, apparently because the difficulty to satisfy conditions of thermodynamically stable compounds with heavy atoms (providing the required large spin-orbit coupling) and inverted conduction and valence bands. Here we demonstrate theoretically the possibility of broadening significantly the material base of TI's by starting from a conventional normal insulator (NI) without band inversion or high-atomic number elements, and realize a NI-to-TI topological phase transition in a single system by applying external electric field. This is predicted for few-layer phosphorene, a novel two-dimensional material that can be isolated through mechanical exfoliation of black phosphorus and unlike graphene and silicene, has a finite band gap. The topological band inversion originates entirely from the field-induced Stark effect. Such tunable phase transition could lead to spin-separated gapless edge states, i.e., quantum spin Hall effect and opens the possibility of making a multi-functional "field effect topological transistor" that could manipulate simultaneously both spins and charge carrier.

Whereas both Topological Insulators (TI) and Normal Insulators (NI) have finite band gap $E_g$ separating the occupied from the unoccupied energy bands, materials that are TI's have, in addition band inversion between certain conduction and valence band states[1], characterized by a negative 'inversion energy' $\Delta_{inv}$, and conveniently diagnosed by the topological invariant quantity[2] $Z_2$ being 1, rather than 0. Consequently, the (*N*-1)-dimensional version (2D surfaces or 1D edges) of such *N*-dimensional bulk TI's (respectively, 3D crystals, or 2D layered crystals) manifest linearly dispersed crossing bands termed Dirac cone having opposite spin currents that are resilient to nonmagnetic chemical passivation and backscattering[2, 3]. Whereas most studies on TI and NI materials tend to focus on separate systems each having its own, fixed $\Delta_{inv}$ – either negative (a TI with $Z_2$=1) or positive (a NI with $Z_2$= 0) – it would appear interesting as well as practically useful to realize a NI-to-TI topological phase transition *in a single system* by manipulating some type of continuous 'knob'. This might be important for converting materials that are normally NI's to TI's, thus expanding the rather limited material base of TI's. Identifying such a NI-to-TI switch would be in particular interesting in 2D layered systems for establishing the device-relevant NI/TI *switchability*. At present *laboratory synthesized* 2D materials that are confirmed TIs include only graphene[4] and silicene[5] having, however, symmetry-enforced zero band gaps if spin-orbit coupling (SOC) is neglected, and extremely small band gaps of $E_g = 10^{-3}$ meV and 1.55 meV, respectively when the small SOC akin to such light elements is considered. Theoretically proposed but not laboratory realized hypothetical 2D TIs include 'Stanene'[6], BiH[7], ZrTe$_5$, HfTe$_5$[8] and Bi$_4$Br$_4$[9] that promise larger band gaps. However, it is not known if the structure assumed in these calculations to produce band inversion is thermodynamically stable or laboratory realizable[10].

One class of approaches to control $\Delta_{inv}$ in a single material family involves utilization of *structural or compositional degrees of freedom* such as alloying a NI with a TI components[11, 12], or exerting quantum confinement on a NI quantum well by cladding it with a suitably chosen barrier material[13, 14, 15, 16], or applying strain[17, 18, 19, 20, 21] to a NI material attempting to induce thereby band inversion. However, controlling $\Delta_{inv}$ via such *material-intrinsic*, structural or compositional knobs have limited flexibility as these approaches generally require different samples (e.g., different alloy compositions or different quantum well thicknesses) and thus (with the exception of strain) represent essentially discontinuous tuning. Such limitations would not be present if the control of $\Delta_{inv}$ and the ensuing topological invariant $Z_2$ would involve a *continuous external knob* applied to a single, fixed material. An example could be applying an external electric field to a 2D NI material such that it transforms electronically into a TI. Such an integration of electric field engineering and topological quantum phases could effectively combine spin and charge transport under an electric field, and thus could enable interesting physics and potential applications[22]. However, although electrically-controlled *insulator-to-metal* transition[23, 24, 25] has been amply demonstrated in 2D systems (underlying the CMOS technology; proposed for spintronics), achievement of electrically-controlled *topological phase transitions* in 2D systems is rare[26]. Recently such approach has been theoretically proposed for $Sb_2Te_3$ that is a TI in its bulk form (structure type *R-3mH*). For thin layers containing up to 4 monolayers of $Sb_2Te_3$ segments (quintuple layer), DFT calculations of Kim et al. predicted a transition from an NI to a TI under electric field[26]. Subsequent scanning tunneling microscopy (STM) measurement of surface state spectrum by Zhang et al. has verified the band-gap tuning under low field, indicating the possibility for such topological phase transition[27]. Note that $Sb_2Te_3$ is a high atomic-number material that is already a TI in bulk form (but not as thin film), so in this case the electric field effect did not change the fundamental topological character of this compound.

Phosphorene is laboratory realized 2D material that can be made at various monolayer thicknesses by mechanical exfoliation of black phosphorus[28, 29]. The latter is a 3D elemental solid, a layered allotrope of white and red phosphorus that is thermodynamically stable in bulk form at room temperature and pressure. Phosphorene is a normal insulator with a sizable band gap of ~1.5 eV for single-layer and positive inversion energy $\Delta_{inv} = E_g$. It is also predicted to have high-mobility at room temperature (~$10^3$ $cm^2V^{-1}s^{-1}$), significant transport anisotropy within the 2D plane and linear dichorism[30, 31, 32, 33]. A field effect transistor (FET) action was indeed demonstrated in a few-layer phosphorene by manipulating the doping level via back-gate voltage[29], leading to current modulation (on-off ratio) of the order of $10^5$. However, unlike other 2D monolayer elemental phases, such as graphene and silicene that are TIs, single-layer as well as few-layer phosphorene (till bulk) is an NI.

Using density functional theory (DFT) with electric field applied self-consistently to the stack of few-layer phosphorene we predict that there will be a transition from a NI (positive $\Delta_{inv}$ and $Z_2 = 0$) to a TI (negative $\Delta_{inv}$ and $Z_2 = 1$) and eventually to a metal. Unlike $Sb_2Te_3$ whose bulk form is already a TI, induced by the SOC of heavy atoms, here black phosphorus is a low atomic-number NI at all thicknesses, so the prediction that this material can be converted to a TI by application of an electric field provides a totally field-induced phase transition and thus holds the potential of expanding the hitherto limited material base of TI by converting NI materials to TI's. The phase sequence is as follows: In the starting normal insulator phase there is a direct band gap at the Brillion zone center Γ separating the unoccupied conduction band C1 from the occupied valence band V1. This insulating gap decreases monotonically as the electric field is applied. Beyond a critical field, which gets reduced as the stack gets thicker, a band inversion occurs whereby C1 becomes occupied and V1 becomes unoccupied and the calculated topological invariant becomes $Z_2 = 1$. Including the SOC, the Dirac-like band crossing becomes gapped, rendering the system a 2D TI with an occupied vs. unoccupied band gap of 5 meV. Finally, at higher fields band C1 off Γ moves further down in energy and touches the Fermi level, and thus the 2D system becomes metallic while retaining the topological non-trivial feature. Such flexibly tuned phase transition would lead to spin-separated gapless edge states, i.e., quantum spin Hall effect, as well as normal insulating and conducting states. This prediction, building on the current FET technology, opens the possibility to make a *field effect topological transistor* that could manipulate both spin and charge carrier simultaneously.

***Crystal structure and zero field electronic structure of few-layer phosphorene.*** Bulk black phosphorus (space group *Cmca*, $D_{2h}^{18}$) is a layered material in which individual atomic layers are bonded together by van der Waals interactions. Each P atom forms within its layer *$sp^3$*-like hybridization by covalently bonding with the three adjacent P atoms, forming a buckled honeycomb structure, as shown in Figure 1a and b. Each monolayer consists of two P sublayers, and thus the three P-P bonds can be classified as having two in-plane bonds within a sublayer plus an out-of-plane bond with another sublayer. Bulk 3D black phosphorus consisting of an infinite number of monolayers has lower DFT energy per phosphorene layer compared with finite layer stacks. Indeed the "peeling energy" for an *n*-layer structure, i.e., the energy change in going from 3D bulk to a finite *n*-layer structure $D(n) = \frac{E(N)-E(N-n)-E(n)}{n}|_{N\to\infty}$, is around -60 meV/atom for *n* = 1 - 4, similar to the calculated counterpart of monolayer $MoS_2$ -73 meV/atom. This order of magnitude implies rather weak interlayer coupling[34] allowing practical peeling.

The band gap of bulk BP (measured 0.3 eV[35]) increases monotonically with decreasing number of layers *n*. Our DFT-PBE calculation (see Methods Section) reveals that the band gaps of *n* = 4, 3, 2 and 1 layers of phosphorene are all direct at $\Gamma$ with $E_g$ = 0.07, 0.19, 0.42 and 0.87 eV, respectively, in agreement with previous calculations based on the same method[30, 36]. Since the PBE exchange correlation often underestimates the band gap, we have also used the more accurate hybrid functional (HSE06). The HSE06 band gap under zero field is 0.65 eV, larger than the PBE result of 0.3 eV, and in agreement with previous calculations on the same level[30, 36]. As a result, the critical field increases from 0.3 V/Å to 0.7 V/Å. On the other hand, the decreasing trend of band gap versus electric field remains unchanged in HSE. Furthermore, the band symmetry of V1 and C1 only depend on the critical field, but not the exchange functional we choose, indicating that the NI-TI transition is still robust but requires larger field (more details given later).

In what follows we use *n* = 4 layer phosphorene at PBE level. Figure 2a shows at zero field a normal insulator with a direct band gap located at $\Gamma$ point. Both bands C1 and V1 exhibit $sp^3$ character of P atom but differ in their symmetric representations at the $\Gamma$ point (point group $D_{2h}$): band C1 (blue) has the representation $A_g$ ($\Gamma_1$), whereas band V1 (red) has the representation $B_{3u}$ ($\Gamma_8$) that is anti-symmetric with respect to the operators including inversion, $C_{2y}$ and $C_{2z}$ of two-fold rotation axis. In addition, the band dispersion along the $k_x$ and $k_y$ direction shows strong anisotropy (see the 2D rectangular Brillouin zone in Figure 1c). The effective mass along $\Gamma$-X is much lower than that along $\Gamma$-Y because of the preference of bonding direction along *x* direction[30], as shown in Figure 1a. The topological invariant at zero field is $Z_2$ = 0, confirming that few-layer phosphorene is a NI.

***From NI to TI: Band gap closure and Band inversion.*** We search for the topological phase transition by applying an external field $F_\perp$ to the *n*-layer stack for $n = 1 - 4$. In order to understand the main effects of the electric field we first exclude SOC in the calculation. Due to Stark effect, the energy shift caused by the difference of electrostatic potential between different layers can be approximately expressed as $\Delta E = -F^* ed$, where $F^*$ is the screened electric field and $d$ is the interlayer distance. In order to monitor the band inversion, we define the inversion energy as $\Delta_{inv} = E_{\Gamma 1c} - E_{\Gamma 8v}$ (see Figure 2a). With $F_\perp$ increasing, the band gap $E_g$ decreases monotonically, and so does $\Delta_{inv}$, until a critical field $F_c = 0.3$ V/Å. At this field band V1 and band C1 touch at $\Gamma$, as shown in Figure 2b. For $F_\perp > F_c$, the band gap $E_g$ remains zero but $\Delta_{inv}$ becomes negative. This band inversion is induced entirely by the field $F_\perp$ rather than by SOC. Similar proposals in which band inversion is induced by factors other than SOC appear in ZeTe$_5$ and HfTe$_5$[8], GaS and GaSe[18], graphene-like materials[4, 5, 6].

The schematic diagram of the evolution of energy level $\Gamma_{1c}$ and $\Gamma_{8v}$, accompanied with the corresponding magnitude of squared wavefunction is shown in Figure 2e. At zero field, the atomic orbitals of $\Gamma_{8v}$ (red, occupied at zero field) are distributed along the out-of-plane bonds of two P sublayers, but are anti-symmetric along the in-plane bonds within each P sublayer. In contrast, atomic orbitals of $\Gamma_{1c}$ (blue, unoccupied at zero field) have nodes in the out-of-plane bonds, while being symmetric along the in-plane bonds. Once the electric field exceeds $F_c$ we find that the $\Gamma_{1c}$ state is now occupied and the $\Gamma_{8v}$ state is empty, indicating a band inversion with negative $\Delta_{inv}$. Although the charge density tends to aggregate at the side layer because of the electrostatic potential and thus inversion symmetry is broken, the rotation symmetry with respect to $C_{2z}$ two-fold axis still differs $\Gamma_{8v}$ and $\Gamma_{1c}$ by group representation (Actually, $\Gamma_{8v}$ state becomes $\Gamma_{2v}$ after applying field because the symmetry of $\Gamma$ is lowered). This band inversion at $\Gamma$ ensures a non-trivial topological invariant $Z_2 = 1$ and thus achieve the topological phase transition tuned by $F_\perp$.

Figure 3a shows the inversion energy $\Delta_{inv}$ at $\Gamma$ of n = 2, 3, and 4-layer phosphorene as a function of the electric field. For *n* = 2 layer, there is no phase transition within our field range (same case for monolayer phosphorene) because the zero-field gap is large. On the other hand, for both *n* = 3 and 4-layer phosphorene, we detect phase transition with the critical field of 0.55 V/Å and 0.3 V/Å, respectively. The result $\Delta_{inv}$ vs. $F_\perp$ of 4-layer phosphorene considering SOC effect is also shown in Figure 3a. We find that the SOC effect in light atom P is weak, so it doesn't affect much the inversion energy at Γ as well as the critical field $F_c$.

The discussion above pertains to band inversion at Γ, not at the other time reversal invariant momenta (TRIM). The quantity that captures the full topological response over the TRIMs is the $Z_2$ invariant. For few-layer phosphorene under electric fields the calculation of $Z_2$ invariant is not as straightforward as that under zero field, because finite field breaks the inversion symmetry, which means the regular method that evaluates the band parity[1] does not apply. Here we use a more general method by Fu and Kane[37]. For solids that have time-reversal symmetry, the time reversal operator matrix relates time reversed wave functions by

$$B_{\alpha\beta}(K_i) = \langle u_{-K_i,\alpha}|\Theta|u_{K_i,\beta}\rangle, \tag{1}$$

where $u_{K,\alpha}$ is the Bloch wave function for occupied band $\alpha$ at TRIM $K$ and $\Theta$ is the time-reversal operator. At TRIM $K_i$ where the Hamiltonian commutes with the time-reversal operator, $B(K_i)$ is antisymmetric. Its Pfaffian, whose square is equal to the determinant, may characterize an antisymmetric matrix. Then $\delta(K_i) = \frac{\sqrt{det[B(K_i)]}}{Pf[B(K_i)]}$ can take ±1 values. The distribution of all values of $\{\delta(K_i)\}$ will be used below to classify different topological types. Therefore, $Z_2$ is defined as

$$(-1)^{Z_2} = \prod_i \delta(K_i) \tag{2}$$

Figure 3b shows the evolution of $Z_2$ invariant of 2, 3, and 4-layer phosphorene with respect to $F_\perp$. We note that for 3 and 4-layer phosphorene, the jumping point of $Z_2$ exactly correlates to the critical field of $\Delta_{inv}$, which means that it is the band inversion at Γ induced by $F_\perp$ causing the topological phase transition.

*Graphene-like Dirac cones.* Ignoring SOC, when the field $F_\perp > F_c$, the band inversion at Γ is accompanied by a Dirac-like band crossing along Γ-Y Figure 2c suggests that along Γ-Y the band crossing induced by V1 and C1 ensures few-layer phosphorene semimetals with two nodes Λ (0, ±$y_D$) at the Fermi level (also see Figure 1c). This band crossing is protected by fractional translation symmetry due to the different band character of the two bands (red and blue in figure). Therefore, they are describable by two $2 \times 2$ Dirac equations, similar to the "massless Dirac fermions" in graphene and silicene without SOC. This band crossing further implies the band inversion on different sides of $\Lambda$, i.e., $\Gamma$ and Y, and confirms that the band inversion is totally induced by the electric field, rather than by SOC. With $F_\perp$ increasing, V1 and C1 continuously shift upwards and downwards, respectively. As a result, the Dirac cone red shifts towards Y point (see Figure 2d). On the other hand, along $\Gamma$-X there is no band crossing because of the absence of fractional translation symmetry along this high-symmetry line, also indicating the anisotropy between *x* and *y* direction.

The SOC in P is too weak to affect the band inversion. However, it lifts the degeneracy at the Dirac point $\Lambda$ and thus opens a gap at the Fermi level, as shown in Figure 3c. Therefore, the electronic behavior of few-layer phosphorene under field is quite similar with graphene-like 2D structures, as "Dirac semimetals" without SOC and 2D TIs with SOC[5, 38]. The band gap of 4-layer phosphorene under 0.45 V/Å is about 5 meV, much larger than that of its 2D elemental counterpart graphene and silicene ($10^{-3}$ meV and 1.55 meV, respectively). The band-gap opening could be interpreted by the evolution of band symmetry. Under zero field, V1 and C1 at $\Lambda$ belong to different single group irreducible representations under fractional translation operation. When electric field and SOC are both applied, the inversion symmetry is removed and the $\Lambda$ symmetry is reduced. As a result, V1 and C1 have the same irreducible representation and thus an anti-crossing rather than crossing. We note that unlike graphene-like materials, in phosphorene inversion symmetry breaking (by applying electric field) is needed to open the SOC-induced gap. This is because compared with graphene, few-layer phosphorene has a non-symmophic space group containing fractional translations that protect the symmetry difference between V1 and C1. In addition, because the inversion symmetry is broken under electric field, there is spin splitting at the non-time-reversal invariant $\Lambda$ point, as shown in Figure 3c (similar to the Rashba effect at the K valley in gated graphene and silicene[39, 40]).

***From TI to (topological) metal: possible implications* on field effect topological transistor.** Above the critical field, increasing the electric field would enhance the reopened gap at Γ (absolute value of $\Delta_{inv}$), keeping the topological non-trivial phase of few-layer phosphorene. As long as few-layer phosphorene in the TI phase, the quantum spin Hall (QSH) effect should be realized[2, 4]. On the other hand, when the electric field reaches another critical field value of $F_M = 0.6$ V/Å, as shown in Figure 2d, we find that another valley of C1 approaches the Fermi level, and thus the system becomes metallic. However, this phase transition does not affect the inverted bands at Γ and thus the topological character. This is verified by our calculation of the $Z_2$ invariant. Therefore, the phase transition sequence of few-layer phosphorene under increased electric field is NI –to TI – to topological metal. Such phase transition provides the possibility to construct a "field effect topological transistor (FETT)" by applying different gate voltages. Here we propose a dual-gated FETT model composed of few-layer phosphorene as the channel, as shown in Figure 4a. Compared with single gate, a dual-gated device can not only tune the perpendicular electric field working on the channel, but also the Fermi level[41]. Figure 4b-d explains the phase transition and the corresponding spin and charge current on the channel. When $F < F_c$, the FETT performs as "OFF" state because the phosphrene channel remains a normal insulator. When $F > F_M$, The FETT is switched to "ON" state, with the charge current flowing through the entire 2D sheet. Finally, as $F$ falls between two critical fields $F_c$ of and $F_M$, a topological non-trivial "QSH" state occurs, with the 2D sheet remaining insulating but net spin current on the edge of the channel. The spin-up and spin-down electrons propagate oppositely, resilient to nonmagnetic chemical passivation (like H termination) and to normal backscattering.

We note that if such a FETT could be switched between "ON" and "OFF" state, i.e., intrinsically metallic or insulating, there must be "QSH" phase between them. As previously discussed, the non-trivial band gap is 5 meV, much larger than that of its 2D elemental counterpart graphene and silicene. However, it still requires a relative low temperature (< 58 K) to realize the QSH effect. It is because when the decreasing $\Delta_{inv}$ approaches $k_BT$ before critical field, electrons begin to transition thermodynamically, and thus the insulator phase will directly jumps to metal phase without passing through TI phase. Therefore, given that the range of electric field that brings few-layer phosphorene to TI is achievable (e.g., 0.3 – 0.6 V/Å for 4-layer phosphorene), we expect the oppositely propagating spin current to be verified in a reasonable experimental setup.

*Manipulating the required critical field.* Since in a more refined, HSE calculation reveals that the critical electric field can be as large as 0.7 V/A for $n = 4$, it is interesting to discuss ways of reducing this value of critical field. In order to realize the topological phase transition one could increase the number of layers $n$ of phosphorene to get a smaller critical field. It is because (i) the band gap decreases with the layer number, approaching the bulk value (~ 0.3 eV). In the proposed FET made by few-layer phosphorene, the thickness of the sample with good quality lies in the range of 5 – 10 nm, corresponding 10 – 20 layers of phosphorene[29]. By extrapolating HSE06 band gaps of such a thickness, we estimate the band gap is within 0.30 – 0.35 eV. (ii) Due to the Stark effect, the response to electric field becomes more significant with thicker layers. In addition, the band edges still locate at Γ for thicker phosphorene slab until bulk state[29], indicating that the TI phase always appears before the metal phase by applying electric field (assuming a rigid band shift). Therefore, considering the compromise of computational cost, we choose PBE calculations that underestimate the critical field and 4 layers whose $E_c$ can be optimized by simply adding the phosphorene layers to schematically prove the topological phase transition, and we expect it to be verified by upcoming experiments.

*Considerations for other 2D materials that could have electrical-induced NI-to-TI topological transition.* Based on the calculations above one could enquire what are the necessary general conditions for materials that would manifest an electric field induced conversion of a 2D NI to a TI. First, one should be able to close the band by applying a reasonable field. Second, such material needs to have a direct band gap located at a TRIM (for the tetragonal Brillouin zone these are Γ, X, Y, and M), because if the band gap were to occur at non-TRIM then either an indirect band gap or direct band gap would make the metal phase occur before the topological band inversion. Considering all the four TRIMs, the degeneracy of valleys should be an odd number. For example, if the direct gap is located at the X point in a square lattice, the band inversion will still lead to

$Z_2 = 0$ because the X valley is doubly degenerate[42]. Third, after the inversion symmetry is broken by the electric field, the band edge *k*-point of V1 and C1 should have different band character (group representation) to make the band inversion non-trivial. Finally, away from the TRIM where band inversion takes place, V1 and C1 should have the same band character to avoid a band crossing and thus keep the gated system an insulator. With respect to these Design Principles, few-layer phosphorene is different that some of the other well known 2D systems such as transition metal dichalcogenides, graphene and its analogs.

***Conclusion and discussion.*** Recently, Zhang *et al.* has proposed and observed an *insulating-to-metallic* transfer curve in phosporene by sweeping the single-gate voltage[29]. Here we add a crucial ingredient, predicting that in few-layer phosphorene there will be a *normal-to-topological phase transition induced purely by applying an electric field and occurring before the system becomes intrinsically metallic*. The topological band inversion originates entirely from the field-induced Stark effect, while the effect of SOC is to open an energy gap at the Dirac-like band crossing, rendering the system a 2D TI. This topological non-trivial feature persists after the system becomes metallic at higher fields. Such tunable phase transition could lead to spin-separated gapless edge states, i.e., quantum spin Hall effect, which is easy to detect as well as normal insulating and conducting states based on the current FET technology. This finding opens the possibility of making a multi-functional "field effect topological transistor" that could manipulate simultaneously both spins and charge carrier.

*Methods*

The calculations were performed with the Vienna *ab initio* package (VASP)[43]. The geometrical and electronic structures are calculated by the projector-augmented wave (PAW) pseudopotential[44] and the generalized gradient approximation of Perdew, Burke and Ernzerhof (PBE)[45] to the exchange-correlation functional unless specified. Electronic structures calculated by hybrid functional (HSE06)[46] is also provided for comparison Spin-orbit coupling is calculated by a perturbation $\sum_{i,l,m} V_l^{SO} \vec{L} \cdot \vec{S} |l,m\rangle_{i\,i}\langle l,m|$ to the pseudopotential, where $|l,m\rangle_i$ is the angular momentum eigenstate of *i*th atomic site[47].

The plane wave energy cutoff is set to 550 eV, and the electronic energy minimization was performed with a tolerance of $10^{-4}$ eV. All the lattice parameters and atomic positions were fully relaxed with a tolerance of $10^{-3}$ eV/Å. The van der Waals interaction is considered by a dispersion-corrected PBE-D2 method[48]. The vacuum separation in the slab supercell is 20 Å to avoid the interaction between periodic images.

**Acknowledgements**

This work was supported by NSF Grant titled "Theory-Guided Experimental Search of Designed Topological Insulators and Band-Inverted Insulators" (No. DMREF-13-34170).

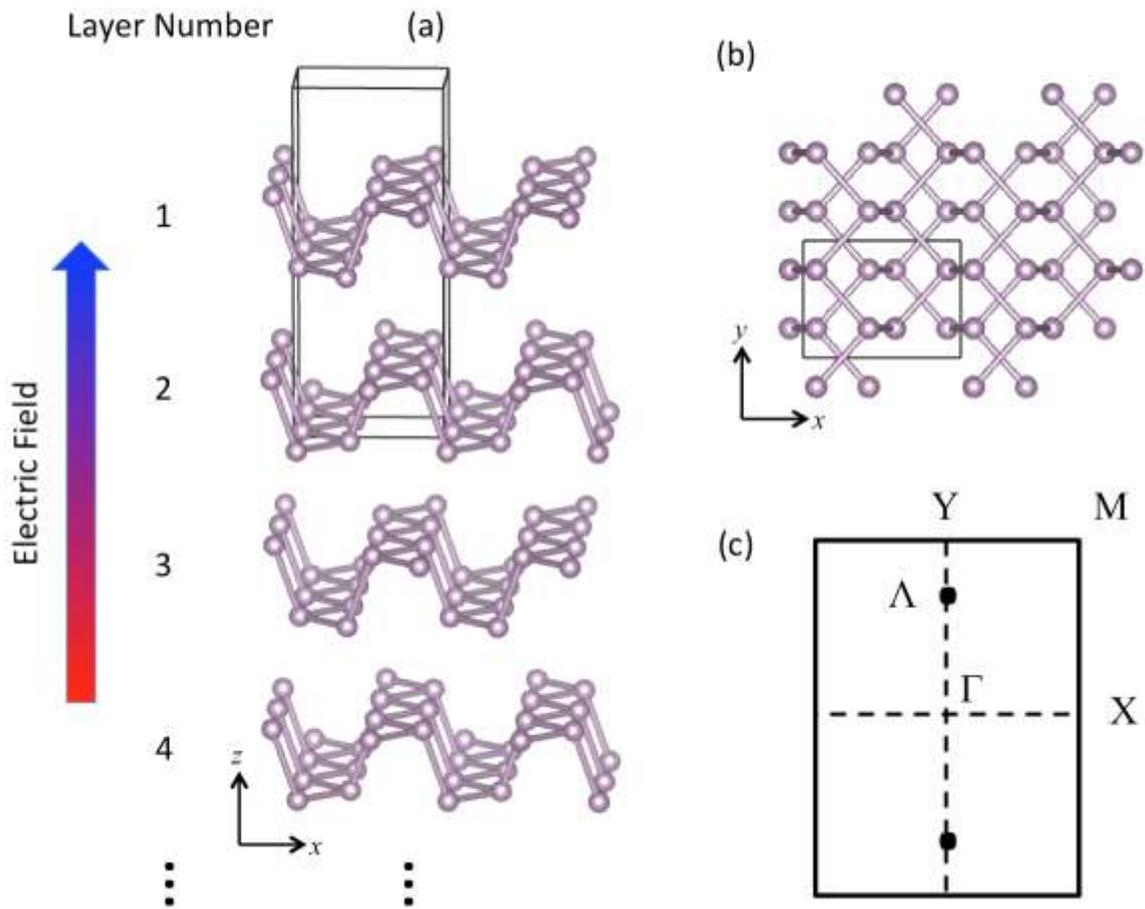

Figure 1: (a) Side view and (b) top view of the crystal structure of multi-layer stack of phosphorene. The black frame indicates the unit cell of bulk black phosphorus. The electric field is applied along the stacking direction. (c) The 2D first Brillouin zone.

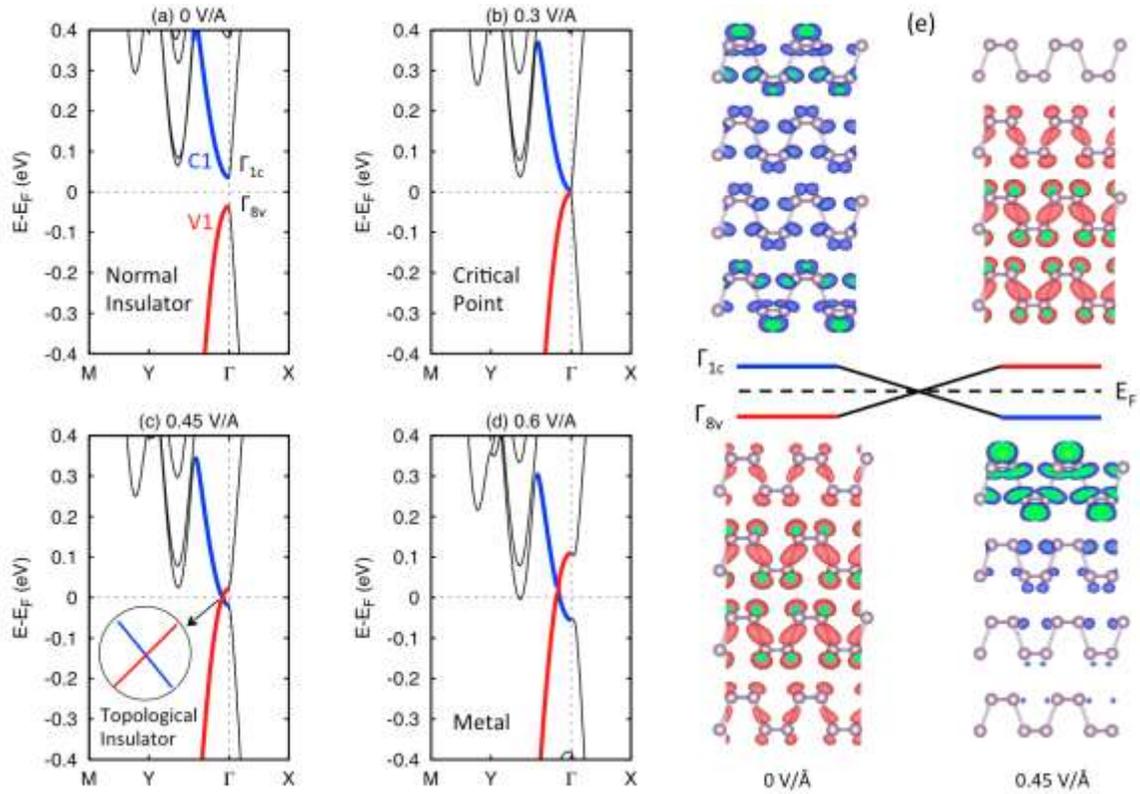

Figure 2: (a) – (d) Band structures of 4-layer phosphorene with an external electric field of (a) 0 V/Å, (b) 0.3 V/Å, (c) 0.45 V/Å, and (d) 0.6 V/Å on PBE level. SOC is not included. The V1 and C1 states are highlighted by red and blue, respectively. (e) Band evolution of $\Gamma_{1c}$ and $\Gamma_{8v}$ before (0 V/Å) and after (0.45 V/Å) electric field with the corresponding magnitude of squared wavefunction of 4-layer phosphorene. Purple balls denote the phosphorus atoms. The amplitude of isosurface is 0.03 a.u.

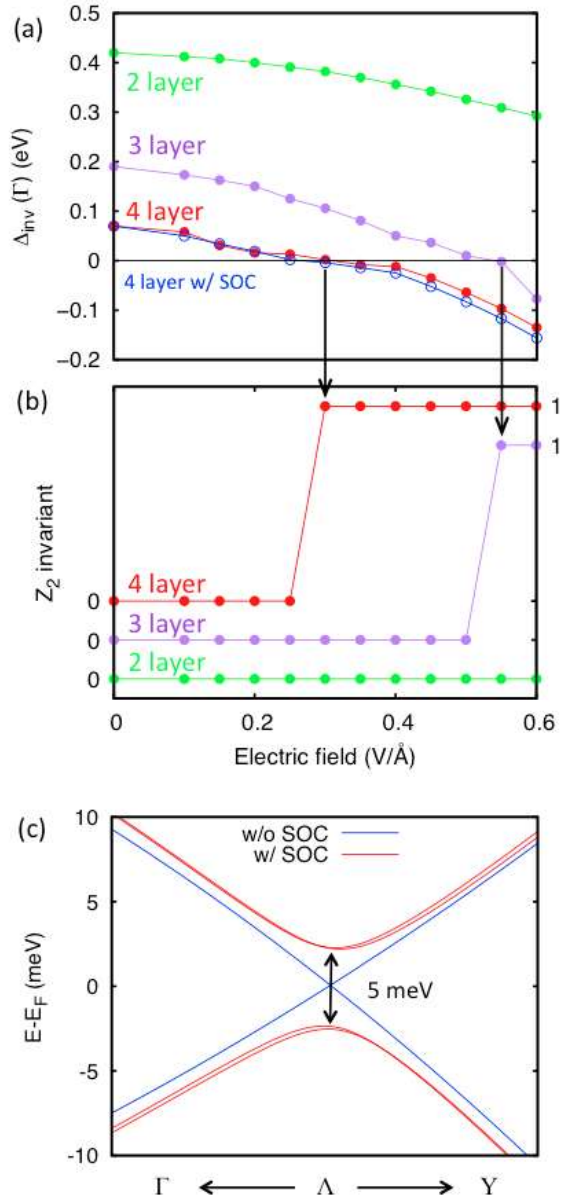

Figure 3: (a) Inversion energy $\Delta_{inv}$ (PBE level) of 2, 3, and 4-layer phosphorene at $\Gamma$ point as a function of the electric field. The negative values indicate non-trivial topological phases. SOC is not included unless specified in 4-layer phorsphorene. (b) $Z_2$ invariant of 2, 3, and 4-layer phosphorene. SOC is included throughout the calculation. It is shown in both panels that for 3 and 4-layer phosphorene, the critical field for the phase transition is 0.55 V/Å and 0.3 V/Å, respectively. (c) Graphene-like band structure of 4-layer phosphorene in the vicinity of $\Lambda$ point with (red) and without (blue) SOC.

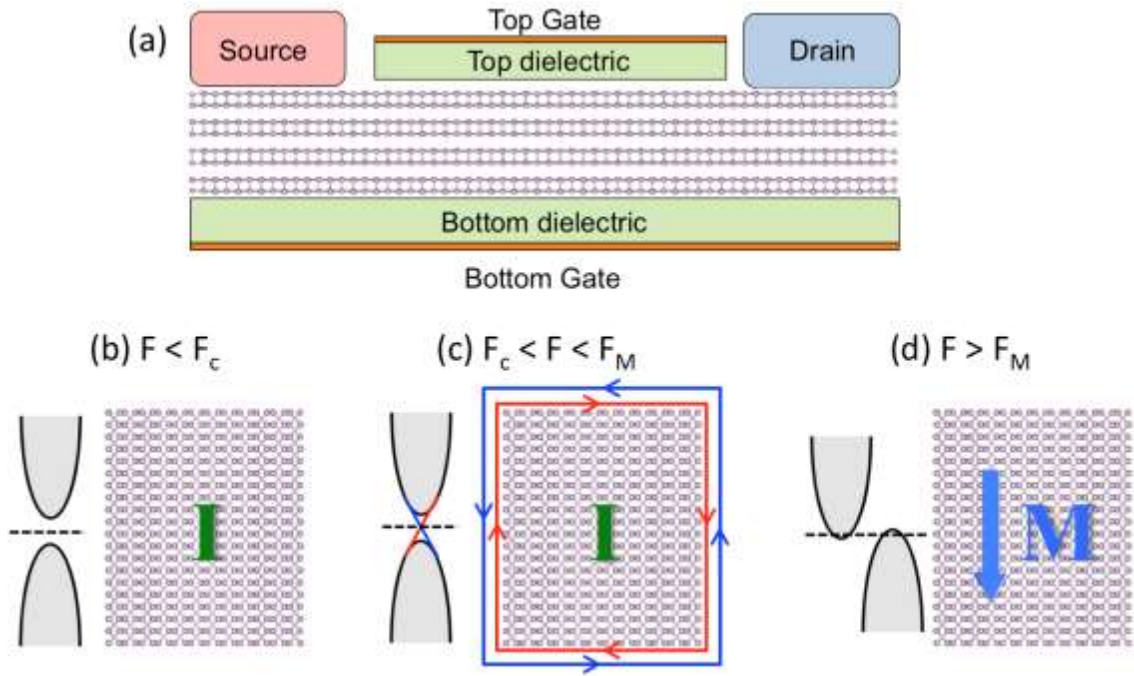

Figure. 4: (a) Model of a dual-gated topological field effect transistor based on a few-layer phosphorene channel. (b-d) schematic band structure and spin and charge current in the phosphorene channel for (b) $F < F_c$, (c) $F_c < F < F_M$ and (d) $F > F_M$. The letter "I" and "M" stands for insulating and metallic, respectively.